\documentclass[aps,prb,twocolumn]{revtex4}
\usepackage{amsmath}
\usepackage{amssymb}
\usepackage{graphicx}

\begin{document}

\preprint{This line only printed with preprint option}

\title{Edge states and topological orders in the spin liquid phases of star lattice}

\author{Guang-Yao Huang}

\author{Shi-Dong Liang}
\email{stslsd@mail.sysu.edu.cn}

\author{Dao-Xin Yao}
\email{yaodaox@mail.sysu.edu.cn}

\affiliation{State Key Laboratory of Optoelectronic Material and
Technology, Guangdong Province Key Laboratory of Display Material and Technology,
School of Physics and Engineering, Sun Yat-sen
University, Guangzhou, 510275, People's Republic of China}
\begin{abstract}
A group of novel materials can be mapped to the star lattice, which
exhibits some novel physical properties. We give the bulk-edge
correspondence theory of the star lattice and study the edge states
and their topological orders in different spin liquid phases. The
bulk and edge-state energy structures and Chern number depend on the
spin liquid phases and hopping parameters because the local
spontaneous magnetic flux in the spin liquid phase breaks the time
reversal and space inversion symmetries. We give the characteristics
of bulk and edge energy structures and their corresponding Chern
numbers in the uniform, nematic and chiral spin liquids. In
particular, we obtain analytically the phase diagram of the
topological orders for the chiral spin liquid states
$SL[\phi,\phi,-2\phi]$, where $\phi$ is the magnetic flux in two
triangles and a dodecagon in the unit cell. Moreover, we find the
topological invariance for the spin liquid phases,
$SL[\phi_{1},\phi_{2},-(\phi_{1}+\phi_{2})]$ and
$SL[\phi_{2},\phi_{1},-(\phi_{1}+\phi_{2})]$. The results reveal the
relationship between the energy-band and edge-state structures and
their topological orders of the star lattice.
\end{abstract}

\maketitle

\section{Introduction}
The discovery of the Integer Quantum Hall effect (IQHE) stimulates
novel fundamental concepts in condensed matter physics, such as the
gauge invariance,\cite{prb23.5632}, edge state,\cite{prb25.2185} and
Chern number\cite{prl49.405}. In particular, Y. Hatsugai reveals the
relationship between Chern number and edge states in the
IQHE,\cite{prl71.3697,prb48.11851} which provides another way from
the edge state to understand the topological order in finite
systems.\cite{njp11.123014} Moreover, the successful synthesis of
nano and novel materials, such as the triangular organic material
$\kappa-BEDT(CN)_{3}$,\cite{prl91.107001} the kagome lattice
herbertsmithite\cite{jacs127.13462} and the three-dimensional
hyperkagome lattice magnet $(Na_{4}Ir_{3}O_{8})$,\cite{prl99.137202}
provide many opportunities to examine theoretically and
experimentally some novel physical properties, including the
topological properties,\cite{prl71.3697,prb48.11851} fractionalized
excitation\cite{prl101.197202,prl101.197201}, singlet valence-bond
solid states,\cite{jap69.5962,prb68.214415,prb76.180407}, and edge
states. Theoretically, these materials can be mapped to novel
geometric lattices, such as the star lattice which is also called
the triangle-honeycomb lattice\cite{ap321.2,prl99.247203}, Fisher
lattice or decorated honeycomb lattice.\cite{prb81.104429} These
geometric lattice models provide a new view to understand the
geometric effect and spin
frustration.\cite{prb81.134418,prb62.R6065,njp11.123014} In
particular, the spin models on the star and honeycomb lattices have
shown many novel phases including the Abelian and non-Abelian
anyons, chiral spin-liquid phases, topological orders,
\cite{ap321.2,prl99.247203,prb81.104429}, magnetic
orders\cite{prb80.064404,prb81.134418}, and topological
insulator.\cite{prb81.205115,prb82.075125}

The IQHE reveals some novel transport properties of electrons in
two-dimensional (2D) systems, in which the edge states play a key
role and the Hall conductance can be expressed in terms of Chern
number.\cite{prb25.2185,prl49.405} Interestingly, the bulk-edge
correspondence discovered by Y. Hatsugai
\cite{prl71.3697,prb48.11851} becomes an efficient method to explore
the edge states in various 2D systems, such as the honeycomb lattice
(Graphene)\cite{prb74.205414} and spin-chiral ferromagnetic kagome
lattice.\cite{prb77.125119,njp11.123014} However, the star lattice
consists of a special lattice geometry, which can be mapped to a
class of materials and cold atoms in optical lattices. The mean
field study of Heisenberg model demonstrates the existence of
several spin liquid phases, which depend on the flux configurations
of two triangles and one dodecagon in the unit
cell.\cite{prb81.134418}  A natural question arises: what is the
relationship between the spin liquid phases and edge states on the
star lattice with boundaries?

\begin{figure}[t]
\includegraphics[width=3.0in]{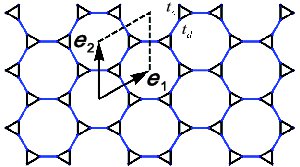}
\caption{(Color online) A star lattice with basis $e_{1}$ and
$e_{2}$. There are six sites in the unit cell. $t_{c}$ represents
the hopping amplitude inside the triangle (the bond in black), and
$t_{d}$ corresponds to the hopping amplitude between different
triangles (the bond in blue).}
\end{figure}

In this paper, we focus on the edge states and their topological
orders on the star lattice with boundaries. We begin with a 2D
tight-binding model with the Hund's rule coupling. It can be mapped to an
effective tight-binding spinless model.\cite{prb62.R6065} The mean
field approach predicts that there exists several spin liquid phases
in the ground states with the local time reversal symmetry
breaking.\cite{prb81.134418} We use the bulk-edge correspondence
method to analyze the edge states and their topological orders on
the star lattice with boundaries.

This paper is organized as follows. In Sec. II, we introduce the
tight-binding model with boundaries and map it to a spinless
tight-binding model. In Sec. III, we give the bulk-edge
correspondence for the star lattice. We present the edge states and
their corresponding Chern numbers in various phases in Sec. IV.
Finally, we give the discussion and conclusions.

\begin{figure}[t]
\includegraphics[width=3.5in]{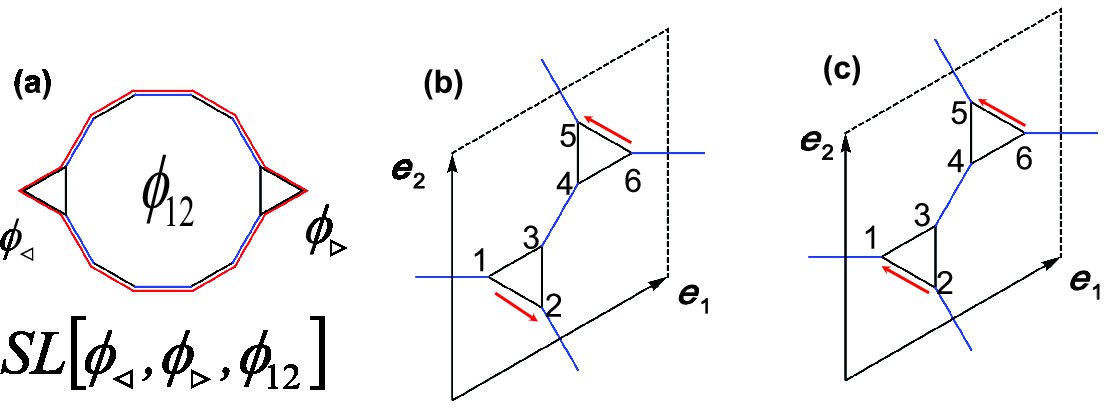}
\caption{(Color online) The elementary plaquette of star lattice
contains two inequivalent triangles $\vartriangleleft$,
$\vartriangleright$, and one dodecagon. The magnetic flux
configurations are labeled by
$SL[\phi_{\vartriangleleft},\phi_{\vartriangleright},\phi_{12}]$
following Ref. \cite{prb81.134418}.}
\end{figure}
\section{Model and spin liquid phases}
In order to understand the relationship between the lattice
geometry, edge states, and their topological properties, we consider
the star lattice with boundaries, in which the conducting electrons
move in a local spin background and couple with them by the Hund's
rule to form a double-exchange system. The corresponding
tight-binding Hamiltonian can be written as,
\begin{equation}
H=\sum_{\langle i,j\rangle \sigma}t_{ij} (c_{j\sigma}^{\dagger}c_{i\sigma}+H.c.)
-J\sum_{i}c_{i\alpha}^{\dagger}\boldsymbol{\sigma}_{\alpha\beta}\cdot\boldsymbol{S}_{i}c_{i\beta}
\label{eq:originalHamilton}
\end{equation}
where $t_{ij}$ is the hopping amplitude between two nearest
neighboring sites $\langle i,j\rangle$;
$c_{i\sigma}^{\dagger}(c_{i\sigma})$ is the creation (annihilation)
operator on site i with spin $\sigma$; $\boldsymbol{S}_{i}$ is the
local spin on site i, which couples with the conducting electron
spins with the effective coupling constant $J$. We consider that the
local spins are approximately classical and the coupling $J$ is
strong enough to have the hoping electrons to align them to the
local spin $S_{i}$ on each site with the spinon function,
$|\chi\rangle=(e^{a_i}\cos(\theta_{i}/2),e^{i(a_{i}+\phi_{i})}\sin(\theta_{i}/2))$,
where $(\theta_{i},\phi_{i})$ are the spinon parameters. In this
spinon representation, the Hamiltonian
Eq.(\ref{eq:originalHamilton}) can be mapped to an effective
tight-binding Hamiltonian,
\begin{equation}
H_{eff}=\sum_{\langle i,j\rangle}(t_{ij}^{eff}c_{i}^{\dagger}c_{j}+H.c.)
\label{effH}
\end{equation}
where the effective hopping amplitude\cite{prb62.R6065}
\begin{eqnarray}
t_{ij}^{eff}&=&t_{ij}\left[\cos(\frac{\theta_{i}}{2})\cos(\frac{\theta_{j}}{2})+e^{-(\phi_{i}-\phi_{j})}\sin(\frac{\theta_{i}}{2})\sin(\frac{\theta_{j}}{2})\right]e^{ia_{ij}}\nonumber\\
 &=&t(\theta_{ij},\phi_{ij})e^{ia_{ij}}
\end{eqnarray}
where the phase $a_{ij}$ is the vector potential generated by spin
and corresponds to the Berry phase felt by the hopping electron. It
is noted that the unit cell of the star lattice contains two
triangular plaquettes and one 12-site dodecagon plaquette. The mean
field study of Heisenberg model on the star lattice has revealed the
existence of several spin liquid phases.\cite{prb81.134418} The spin
liquid phases depend on the flux figuration of the unit cell, which
is labeled by the notation
$SL[\phi_{\vartriangleleft},\phi_{\vartriangleright},\phi_{12}]$.\cite{prb81.134418}
In terms of the original spin variable, the fluxes on the triangular
plaquettes correspond to the scalar spin chiralities
$\boldsymbol{S}_{1}\cdot\boldsymbol{S}_{2}\times\boldsymbol{S}_{3}$,
while $\phi_{12}$ is related to the 12 spins around the dodecagon
loop. In the uniform spin liquid phase, $SL[0,0,0]$,
$t_{ij}^{eff}=t(\theta_{ij},\phi_{ij})\in\mathbb{R}$. For the
non-uniform spin liquid state, the spin chirality arises, the
fermion hopping will acquire a phase
$t_{ij}^{eff}=t(\theta_{ij},\phi_{ij})e^{ia_{ij}}$, rendering a
nonzero flux for a fermion moving around a loop
$\phi=\sum_{loop}a_{ij}$. $t(\theta_{ij},\phi_{ij})$, in principle,
depends on the angles between spins $\boldsymbol{S}_{i}$ and
$\boldsymbol{S}_{j}$. However in the mean field approximation,
$t(\theta_{ij},\phi_{ij})$ should be independent on the
angles,$(\theta_{ij},\phi_{ij})$ for the spin liquid phases because
the fluxes through plaquettes are periodic in the whole
lattice.\cite{prb81.134418} Thus the hopping parameters
$t_{ij}^{eff}$ can be classified into two independent variables on
the star lattice. $t_{c}$ labels the hopping amplitude inside the
triangles and $t_{d}$ is the hopping amplitude between different
triangles. For convenience, we introduce
$r\equiv\frac{t_{d}}{t_{c}}$ to measure the ratio of these two
hopping amplitudes.

For a given flux configuration, $\phi=\sum_{loop}a_{ij}$, there are
different phase configurations $a_{ij}$. The effect of the phase
configuration can shift the whole energy band in the k space, but do
not modify the energy band structure. Namely, different choices of
the phase configurations do not change the edge state and their
topological properties of the star lattice. This allows us to set a
simple choice of the phase configuration for various spin liquid
phases to study their edge states and topological orders.

\begin{figure}[t]
\includegraphics[scale=0.25]{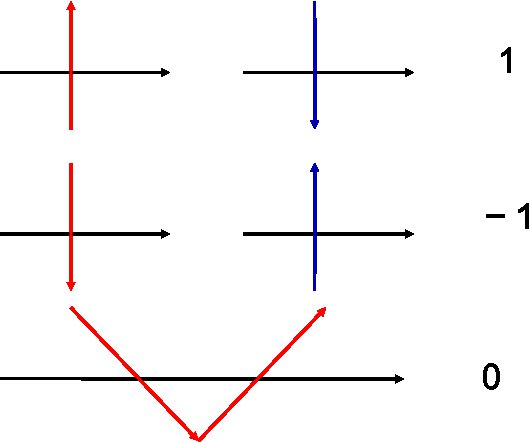}
\caption{(Color online) The intersection number between the
canonical loop on the complex-energy surface (Riemann surface) and
the trace of the edge state energy  Ref. \cite{prb48.11851}.}
\end{figure}

\section{Edge states and topological orders}
\subsection{Bulk-edge conrrespondance theory of star lattice}
In general, we consider a strip of star lattice with the boundary
along the $\boldsymbol{e}_{1}$ direction and the periodic infinite
$\boldsymbol{e}_{2}$ direction shown in Fig.1. We assume that the
spin liquid phase
$SL[\phi_{\vartriangleleft},\phi_{\vartriangleright},\phi_{12}]$,
where the fluxes satisfy the constraint,
$\phi_{\vartriangleleft}+\phi_{\vartriangleright}+\phi_{12}=0$.
Using the Bloch theorem in the $\boldsymbol{e}_{2}$ direction,
$c_{j}=\frac{1}{L_{2}}\sum_{\boldsymbol{k}}e^{i\boldsymbol{k}\cdot\boldsymbol{e}_{2}}c_{n\ell}(k)$,
where $n$ labels the the unit cell and $\ell$ labels the sites in
the unit cell. We set $\boldsymbol{k}\cdot\boldsymbol{e}_{2}=k$ for convenience. The Hamiltonian in Eq.(\ref{effH}) can be written as
\begin{equation}
H_{eff}=\sum_{k}\boldsymbol{C}^{\dagger}(k)\boldsymbol{h}(k)\boldsymbol{C}(k),
\label{effHK}
\end{equation}
where $\boldsymbol{C}^{\dagger}(k)=(c_{1,1}^{\dagger}(k)...c_{1,6}^{\dagger}(k) c_{2,1}^{\dagger}(k)...c_{2,6}^{\dagger}(k)...c_{N_{1},6}^{\dagger}(k))$, and
\begin{equation}
\boldsymbol{h}(k)=\left[\begin{array}{ccccc}
d(k) & v & 0 & \cdots  & 0 \\
v^{\intercal}  & d(k) & v & 0 & \vdots\\
0 & v^{\intercal}  & \ddots   & v & 0\\
\vdots & 0 & v^{\intercal} & d(k) & v\\
0 & \cdots & 0 & v^{\intercal} & d(k)
\end{array}\right]_{N_{1}\times N_{1}}
\label{unitcell}
\end{equation}
with
\begin{eqnarray}
d(k)&=&t_{c}\left[\begin{array}{cccccc}
0 & e^{-i\phi_{1}} & 1 & 0 & 0 & 0\\
e^{i\phi_{1}} & 0 & 1 & 0 & re^{-ik} & 0\\
1 & 1 & 0 & r & 0 & 0\\
0 & 0 & r & 0 & 1 & 1\\
0 & re^{-ik} & 0 & 1 & 0 & e^{i\phi_{2}}\\
0 & 0 & 0 & 1 & e^{-i\phi_{2}} & 0
\end{array}\right]
\label{unitcell}    \\
v&=&\left[
\begin{array}{cccccc}
0 & 0 & 0 & 0 & 0 & 0 \\
0 & 0 & 0 & 0 & 0 & 0 \\
0 & 0 & 0 & 0 & 0 & 0 \\
0 & 0 & 0 & 0 & 0 & 0 \\
0 & 0 & 0 & 0 & 0 & 0 \\
r & 0 & 0 & 0 & 0 & 0%
\end{array}%
\right],
\end{eqnarray}
where $N_{1}$ is the number of unit cell along the $\boldsymbol{e}_{1}$ direction.
The Bloch wave function can be written as
$|\Psi(k)\rangle=\sum_{n,\ell}\psi_{n,\ell}c^{\dagger}_{n,\ell}(k)|0\rangle$, where $n$ runs the unit cell in the $\boldsymbol{e}_{1}$ direction. Inserting it into the Schr\"odinger equation, $H|\Psi\rangle=E|\Psi\rangle$, the solution can be reduced to a set of equations (Harper equation)
\begin{equation}
\left.\left\{ \begin{array}{c}
\psi_{n,2}e^{-i\phi_{1}}+\psi_{n,3}+r\psi_{n-1,6}=\varepsilon\psi_{n,1}\\
\psi_{n,1}e^{i\phi_{1}}+\psi_{n,3}+r\psi_{n,5}e^{-ik}=\varepsilon\psi_{n,2}\\
\psi_{n,1}+\psi_{n,2}+r\psi_{n,4}=\varepsilon\psi_{n,3}\\
r\psi_{n,3}+\psi_{n,5}+\psi_{n,6}=\varepsilon\psi_{n,4}\\
\psi_{n,4}+r\psi_{n,2}e^{-ik}+\psi_{n,6}e^{i\phi_{2}}=\varepsilon\psi_{n,5}\\
\psi_{n,4}+r\psi_{n+1,1}+\psi_{n,5}e^{-i\phi_{2}}=\varepsilon\psi_{n,6}
\end{array}\right.\right.
\label{Harper}
\end{equation}
where $\varepsilon=\frac{E}{t_{c}}$. Rewriting Eq.(\ref{Harper}) to
a matrix form, we can express it in terms of a transfer matrix form,
\begin{equation}
\left(\begin{array}{c}
\psi_{n+1,1}\\
\psi_{n,6}
\end{array}\right)=M(\varepsilon)\left(\begin{array}{c}
\psi_{n,1}\\
\psi_{n-1,6}
\end{array}\right)
\end{equation}
where $M$ is a $2\times2$ matrix and its elements are
\begin{eqnarray*}
M_{11}(\varepsilon)&=&e^{i\frac{k+\phi_1-\phi_2}{2}}\frac{l_4 l_5-r^2 l_1^2}{r^2 l_1 l_2}\\
M_{12}(\varepsilon)&=&-e^{i\frac{k+\phi_1-\phi_2}{2}}    \frac{l_5}{r l_1} \\
M_{21}(\varepsilon)&=& e^{i\frac{k+\phi_1-\phi_2}{2}}    \frac{l_4}{r l_1}\\
M_{22}(\varepsilon)&=& -e^{i\frac{k+\phi_1-\phi_2}{2}}   \frac{l_2}{l_1}
\label{M}
\end{eqnarray*}
where
\begin{eqnarray*}
l_1&=&2[(\varepsilon ^2-r^2)\cos\frac{k+\phi_1-\phi_2}{2}+ \\
&&2\varepsilon\cos\frac{k}{2} \cos\frac{\phi_1+\phi_2}{2}+\cos\frac{k-\phi_1+\phi_2}{2}] \\
l_2&=&1+r^4-2 (1+r^2) \varepsilon ^2+\varepsilon ^4-2 r^2 \cos k\\
l_3&=&\varepsilon(3+2 r^2+r^4-2(2+r^2) \varepsilon ^2+\varepsilon ^4)-2 r^2 \varepsilon\cos k\\
l_4&=&l_3+2 (1-\varepsilon ^2) \cos{\phi_1}-2 r^2 \cos(k+\phi_1)\\
l_5&=&l_3+2 (1-\varepsilon ^2) \cos\phi_2-2 r^2 \cos(k-\phi_2)
\label{P}
\end{eqnarray*}
We assume that the width $L_{1}$ of the star lattice contains an
integer number of the unit cells, we can get the reduced transfer
matrix
\begin{equation}
\left(\begin{array}{c}
\psi_{L_{1}+1,1}\\
\psi_{L_{1},6}
\end{array}\right)=(M(\varepsilon))^{L_{1}}\left(\begin{array}{c}
\psi_{1,1}\\
\psi_{0,6}
\end{array}\right)
\end{equation}
Considering the boundary condition $\psi_{L_{1},6}=\psi_{0,6}=0$,
the edge energy $\varepsilon_{edge}$ satisfies $(M(\varepsilon)^{L_{1}})_{21}=0$.\cite{prl71.3697,prb48.11851}
For $L_{1}\gg 1$, the criterion for edge states follows \cite{prl71.3697,prb48.11851}
\begin{equation}
|(M(\mu_{j}))_{11}|\begin{cases}
<1 & \textrm{edge states localized in site 1}\\
>1 & \textrm{edge states localized in site \ensuremath{L_{x}-1}}\\
=1 & \textrm{coincide with bulk states}
\end{cases}
\end{equation}

The quantum Hall conductance of systems can be expressed in terms of
the Chern number of U(1) bundle over the magnetic Brillouin
zone.\cite{prl49.405} The bulk-edge correspondence theory reveals
that the Chern number $C(\mu_{j})$ is equivalent to the winding
number of the edge state moving around the hole of Riemann surface
(the complex-energy surface), namely the intersection number between
the canonical loop on the Riemann surface and the trace of the edge
state energy $\mu_{j}$.\cite{prl71.3697,prb48.11851,njp11.123014}
(see Fig. 3). Thus, the quantum Hall conductance can be given by the
winding number of the edge states,
$\sigma_{xy}^{edge}=-\frac{e^{2}}{h}C(\mu_{j})$ when the Fermi
energy lies in the $j$th energy gap. Therefore, we can count the
Chern number from the energy spectrum of the system.

The mean field studies of the star lattice with the Hamiltonian of
Eq.(\ref{eq:originalHamilton}) give various spin liquid
phases.\cite{prb81.134418} It is worth studying that the topological
properties of these spin liquid phases.

\begin{figure}[t]
\includegraphics[width=3.5in]{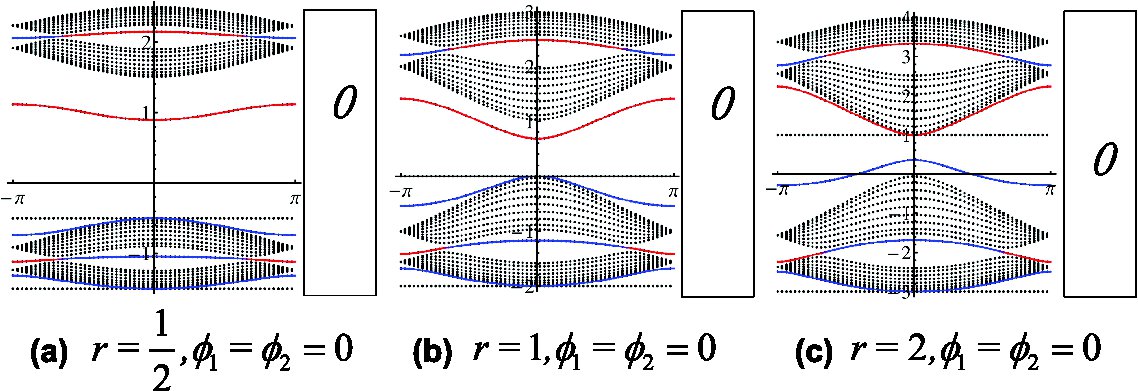}
\caption{(Color online) The energy spectra of $SL[0,0,0]$ for
$r=1/2$ in (a), $r=1$ in (b), and $r=2$ in (c).}
\end{figure}

\subsection{Uniform spin liquid phase: $SL[0,0,0]$}
The uniform spin liquid phase, $SL[0,0,0]$ respects all the
space-group symmetry of the lattice and time reversal symmetry. The
energy spectra of the $SL[0,0,0]$ phase for several set of
parameters $(r,\phi)$ are plotted in Fig. 4. It can be seen that
both of the bulk energy band and edge states are k-symmetric,
$E(k)=E(-k)$, due to the space inverse symmetry, but the edge states
are either embedded in the bulk states or isolated in the gap,
namely there is no nontrivial bulk gap. Interestingly there exist
two flat bands lying in the energy-band gap and touching an
edge-state band, which is caused by interference.
\cite{prb54.R17296} It is similar to a uniform spin liquid on the
kagome lattice and can be spoiled by perturbations, such as the next
nearest neighbor hopping. \cite{prb81.134418} Thus, the Chern number
is not well-defined and corresponds to common metals or insulators
without IQHE. Actually the interactions from spinons can lead to
instability of the uniform spin liquid phase to develop to the
phases with breaking time reversal symmetry.\cite{prb81.134418}

\subsection{Nematic spin liquid phase: $SL[\phi,-\phi,0]$}
The spin liquid phases are characterized by a set of spin chirality
operators.\cite{prb81.134418} Different chiral spin phases exhibit
different local magnetic fluxes. For the nematic spin liquid phase
$SL[\phi,-\phi,0]$, time reversal symmetry is broken
spontaneously,\cite{prb81.134418} we plot the energy spectra for
some parameter $(r,\phi)$ in Fig. 5. It can be seen that the
k-symmetries of both bulk bands and edge states are broken,
$E(k)\neq E(-k)$. It shows that time reversal symmetry breaking
could induce space inverse symmetry breaking that breaks the
k-symmetries of bulk bands and edge states. However, the edge states
are also either embedded in the bulk states or isolated in the gap.
This implies the systems are common metals or insulators, but
without IQHE, as $SL[0,0,0]$ phase.

\begin{figure}[t]
\includegraphics[width=3.5in]{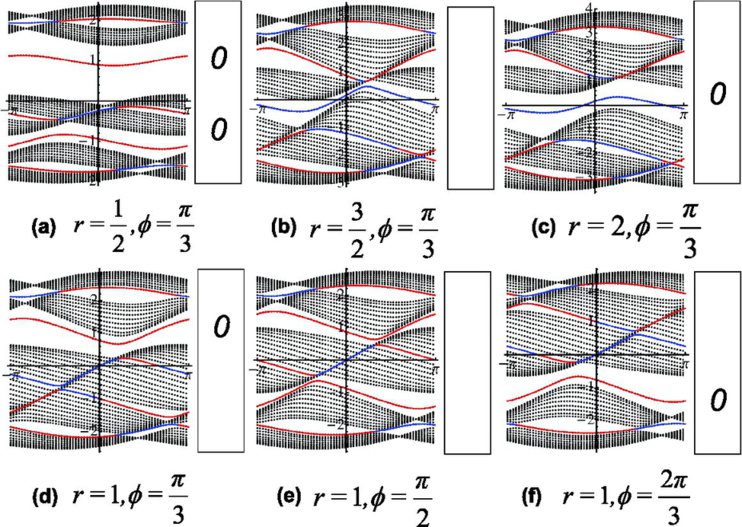}
\caption{(Color online) The energy spectra of $SL[-\phi,\phi,0]$ for
different parameters $(r,\phi)$.}
\end{figure}

\subsection{Chiral spin liquid phase I: $SL[\phi,\phi,-2\phi]$ }
For the chiral spin liquid phase I $SL[\phi,\phi,-2\phi]$, time
reversal symmetry is also broken spontaneously. \cite{prb81.134418}
In principle, the Chern number in the $j$th gap between the bulk
energy bands depends on the parameters $r$ and $\phi$,
$C_{j}(r,\phi)$ (here we use this symbol for Chern number). However,
we find from numerical investigations that the Chern number in the
range of $r$ and $\phi$ obeys the following symmetries :

(1) $C_{j}(r,2\pi-\phi)=-C_{j}(r,\phi)$ for $\phi\in(0,\pi)$;

(2) $C_{j}(r,\pi+\phi)=-C_{5-j}(r,\phi)$ for $\phi\in(0,\pi)$;

(3) $C_{j}(r,\phi)=C_{j}(-r,\phi)$ for $\phi\in(0,2\pi)$;

Thus, we can restrict the parameters only in the range of
$\phi\in(0,\frac{\pi}{2})$ and $r>0$. In Fig. 6 we plot the energy
spectrum for some typical parameters $(t,\phi)$ and $L_{1}=10$. The
Chern number can be counted by the winding number of the torus
formed by two Riemann surfaces.\cite{prl71.3697} (see Fig. 3). The
numbers in the right-hand side of each figures in Figs. 6 are the
Chern number of the system when the Fermi energy lies in the
corresponding energy gap. It can be seen that the bulk energy
spectrum is k-symmetric,  $E(k)=E(-k)$ even though $\phi\neq 0$, but
the edge state has no k-symmetry. It implies that the space
inversion symmetry still holds due to the bulk band k-symmetry, but
the spontaneous time reversal breaking breaks the edge state
k-symmetry.
\begin{figure}[t]
\includegraphics[width=3.5in]{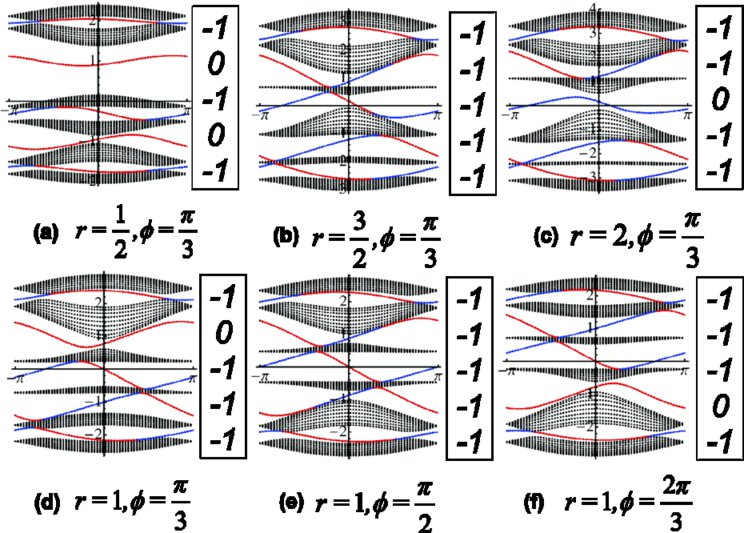}
\caption{(Color online) The energy spectra of $SL[\phi,\phi,-2\phi]$
for different parameters $(r,\phi)$.}
\end{figure}
Different Chern number implies different topological orders of the
system. In order to give the phase diagram in the parameter space,
we try to find out the critical lines in the parameter space. When
the energy gaps close, the Chern number must change, namely a phase
transition happens. Notice that the bulk spectrum has the Krammers
degeneracy, $E(k)=E(-k)$, it implies that the close of the energy
gap of the bulk band happens only at the $\Gamma$ point ($k=0$).
Thus, we solve the eigenenergies of the unit cell Hamiltonian (bulk
band) at $k=0$. Numerical investigation indicates the bulk energy
bands at $k=0$ are linear with $r$, which allows us suppose that the
eigenenergies of the bulk energy band at $k=0$ have the form
$E_{i}(k=0)=\pm r+b_{i}(\phi)$, where $i=1,2,3$. Substituting this
form into the eigen equation of the Hamiltonian of the
two-dimensional star lattice in the k space we can find that
$b_{i}(\phi)$ satisfies a cubic equation,
\begin{equation}
b^{3}-3b-2\cos\phi=0
\end{equation}
When we consider $\phi\in(0,\frac{\pi}{2})$, the solutions of the above cubic equation are
\begin{equation}
\left\{ \begin{array}{c}
b_{1}=2\cos\frac{\phi+2\pi}{3}\\
b_{2}=-2\cos\frac{\phi+\pi}{3}\\
b_{3}=2\cos\frac{\phi}{3}
\end{array}\right.
\end{equation}
which we set $b_{1}<b_{2}<b_{3}$. The closes of the bulk energy band gaps yield the critical lines in the parameter space,
\begin{equation}
\left\{ \begin{array}{c}
r_{c,1}=\sqrt{3}\sin\frac{\phi}{3}\\
r_{c,2}=\sqrt{3}\sin\frac{\pi-\phi}{3}\\
r_{c,3}=\sqrt{3}\sin\frac{\phi+\pi}{3}
\end{array}\right.
\label{critical.line}
\end{equation}
These three functions separate the parameter space $r-\phi$ into
several different regions which have different topological orders
shown in Fig. 7. Different regions in the phase diagram represent
the ground states with different Chern number configurations in
which the Chern number in different positions corresponds to the
Fermi energy in different gaps like Fig. 5. The phases with
different Chern numbers reflect different integer quantum Hall
conductance of the system.

\begin{figure}[t]
\includegraphics[scale=0.5]{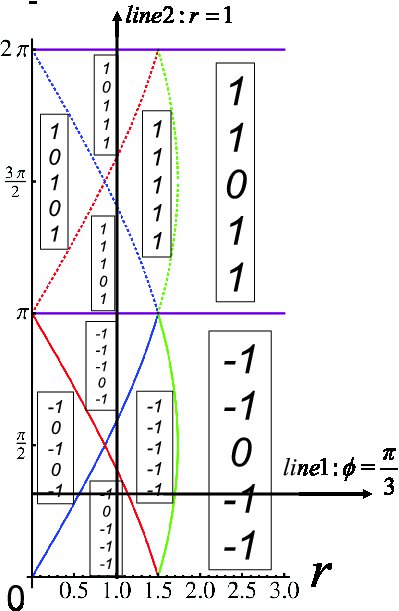}
\caption{The phase diagram of $SL[\phi,\phi,-2\phi]$. Different
region represents different Chern numbers for various filling
fraction, which are even functions of $r$ and periodic functions of
$\phi$ with periodicity $2\pi$. The energy spectrum along line 1 and
line 2 are shown along
(a)$\rightarrow$(d)$\rightarrow$(b)$\rightarrow$(c) and
(d)$\rightarrow$(e)$\rightarrow$(f) in Fig. 6 respectively. }
\end{figure}

\subsection{Chiral spin liquid phase II: $SL[\phi_{1},\phi_{2},-(\phi_{1}+\phi_{2})]$ }
For more general chiral spin liquid phases II
$SL[\phi_{1},\phi_{2},-(\phi_{1}+\phi_{2})]$, the time reversal and
space inverse symmetries are broken spontaneously. For example, the
energy spectra of the spin liquid phase
$SL[\frac{\phi}{3},\frac{2\phi}{3},-\phi]$ for some specific
parameters are shown in Fig. 8. It can be seen that the k-symmetries
of both bulk bands and edge state are broken $E(k)\neq E(-k)$, and
the Chern numbers in some energy gaps are not well-defined, such as
the top three gaps for $r=1,\phi_{1}=\frac{2\pi}{3}$ and
$\phi_{2}=\frac{4\pi}{3}$. Because the degenerate points of the bulk
band and edge state are not at the $k=0$ point, the phase transition
lines can not be solved easily.

\begin{table}[b]
\caption{The energy band properties of different spin liquid phases}
\label{table1} \centering
\begin{tabular}{lcccccccc}\hline\hline
   &  EKS & BKS & TRS & SIS & & \multicolumn{3}{c}{Chern number} \\\hline
   &      &     &     &     & $r$: &\ \ \ \ $\frac{1}{2}$\ \ \ \ &\ \ \ \ $1$\ \ \ \ & $2$    \\\hline
$SL[0,0,0]$ & yes & yes &yes & yes & & $-1$ & $-1$ & x \\
$SL[\frac{\pi}{3},-\frac{\pi}{3},0]$ & no & no &no & no & & x & x & $0$ \\
$SL[\frac{\pi}{3},\frac{\pi}{3},-\frac{2\pi}{3}]$ & no & yes & no & yes & & $-1$ & $-1$ & $0$\\
$SL[\frac{\pi}{3},\frac{2\pi}{3},-\pi]$ & no & no & no & no & & $0$
& $-1$ & $0$ \\\hline\hline
\multicolumn{9}{l}{EKS: Edge-state k-symmetry; BKS: Bulk band k-symmetry;}\\
\multicolumn{9}{l}{TRS: Time reversal symmetry; SIS: Space inverse symmetry;}\\
\multicolumn{9}{l}{x: non-well-defined Chern number.}
\end{tabular}
\end{table}

\section{Discussions}
To compare the basic energy-band properties of different spin liquid
phases, we assume that the Fermi energy lies in the middle of the
middle energy-band gap. The energy band symmetry and Chern numbers
for some specific cases are listed in Table I.

It can be seen from Table I that for $SL[0,0,0]$ both of the edge
k-symmetry (EKS) and bulk k-symmetry (BKS) are held due to the time
reversal invariance and space inversion invariance. When $r>2$, the
Chern number becomes non-well-defined. The cases in the $2\sim4$ lines in
Table I indicate that there is no EKS, but BKS remains for
$SL[\frac{\pi}{3},\frac{\pi}{3},-\frac{2\pi}{3}]$. This implies that
the spontaneous time reversal symmetry breaking does not always
induce the space inversion symmetry breaking and the EKS can be
broken by the time reversal symmetry.

The results in Table I reveal that the Chern number depends on not
only the time reversal and space inverse symmetries, but also the
parameters $(r,\phi_{1},\phi_{2},\phi_{12})$ of the star lattice.
The ground states become normal metals or semiconductors for the
phases without the well-defined Chern number. Interestingly, there
is a topological invariance for the exchange of the magnetic fluxes
in the two triangles $SL[\phi_{1},\phi_{2},-(\phi_{1}+\phi_{2})]$
and $SL[\phi_{2},\phi_{1},-(\phi_{1}+\phi_{2})]$. They have the same
Chern numbers and their bulk energy bands are k-asymmetric. These
findings indicate some new phases in the 2-dimensional
materials.\cite{Nagaosa}

\begin{figure}
\includegraphics[width=3.3in]{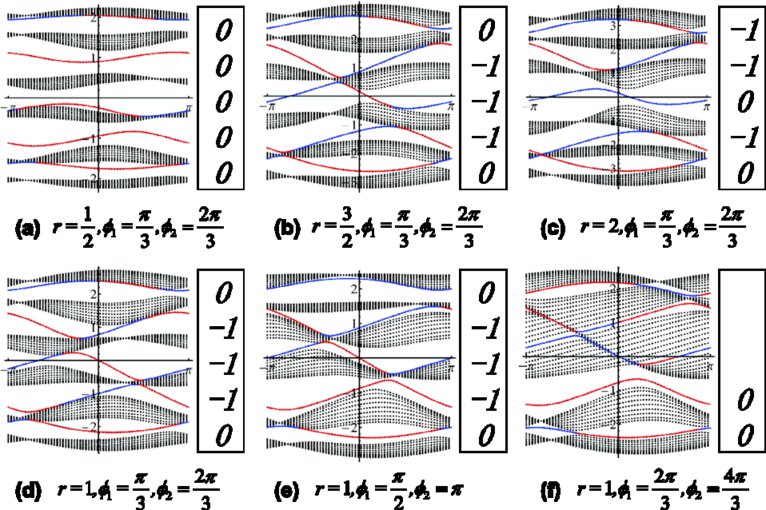}[t]
\caption{(Color online) The energy spectra of
$SL[\phi_{1},\phi_{2},-(\phi_{1}+\phi_{2})]$ for different
parameters $(r,\phi)$.}
\end{figure}

\section{Conclusions}
In summary, we have studied the edge states and their topological
orders in the different spin liquid phases of star lattice by using
the bulk-edge correspondence theory. The bulk and edge-state energy
structures and Chern number depend on the spin liquid phases and
hopping parameters because the local spontaneous magnetic flux in
the spin liquid phases breaks the time reversal and space inversion
symmetries. We have given the characteristics of bulk and edge
energy structures and their corresponding Chern numbers in the
uniform, nematic and chiral spin liquid phases. In particular, we
have obtained analytically the phase transition lines of different
topological phases and their corresponding phase diagrams for the
chiral spin liquid states $SL[\phi,\phi,-2\phi]$. We have also found
that the topological invariance for the spin liquid
phases,$SL[\phi_{1},\phi_{2},-(\phi_{1}+\phi_{2})]$ and
$SL[\phi_{2},\phi_{1},-(\phi_{1}+\phi_{2})]$. The results tell us
the relationship between the energy-band and edge-state structures
and their topological orders of the star lattice. Especially, this
star lattice has been synthesized in the material called iron
acetate recently.\cite{acie46.6076} Therefore, our results provide a
Hall conductance experimental guideline to discriminate the spin
liquid phases in real materials and cold atoms in optical lattice.
The changes of filling fraction could be implemented by tuning the
applied gate voltage. These results can also give some hints for
understanding the Heisenberg model on the star lattice.

\begin{acknowledgments}
We thank Ming-Liang Tong and Xiao-Ming Chen for helpful discussions.
G.-Y. Huang thanks An Zhao and Jie-Sen Li for useful discussions on
numerical calculations. This work is supported by the Fundamental
Research Funds for the Central Universities of China (11lgjc12 and
10lgzd09), NSFC-11074310, MOST of China 973 program (2012CB821400),
Specialized Research Fund for the Doctoral Program of Higher
Education (20110171110026), and NCET-11-0547.

\end{acknowledgments}


\begin{thebibliography}{References}

\bibitem{prb23.5632}R. B. Laughlin, Phys. Rev. B 23, 5632(1981)
\bibitem{prb25.2185}B. I. Halperin, Phys. Rev. B 25, 2185 (1982)
\bibitem{prl49.405}D. J. Thouless, M. Kohmoto, M. P. Nightingale and M. den Nijs, Phys. Rev. Lett. 49, 405 (1982)
\bibitem{prl71.3697}Y. Hatsugai, Phys. Rev. Lett. 71, 3697 (1993).
\bibitem{prb48.11851}Y. Hatsugai, Phys.Rev. B 48, 11851 (1993).
\bibitem{njp11.123014}Zhigang Wang, Ping Zhang, New J. Phys. 11, 123014 (2010)
\bibitem{prl91.107001}Y. Shimizu, K. Miyagawa, K. Kanoda, M. Maesato and G. Saito, Phys. Rev. Lett. 91, 107001 (2003)
\bibitem{jacs127.13462}Matthew P. Shores, Emily A. Nytko, Bart M. Bartlett and Daniel G. Nocera, J. Am. Chem. Soc. 127, 13462(2005)
\bibitem{prl99.137202}M. Sasaki, K. Hukushima, H. Yoshino and H. Takayama, Phys. Rev. Lett. 99, 137202 (2007)
\bibitem{prl101.197201}Yi Zhou, Patrick A. Lee, Tai-Kai Ng and Fu-Chun Zhang, Phys. Rev. Lett. 101, 197201 (2008)
\bibitem{prl101.197202}Michael J. Lawler, Arun Paramekanti, Yong Baek Kim and Leon Balents, Phys. Rev. Lett. 101, 197202 (2008)
\bibitem{prb68.214415}P. Nikolic and T. Senthil, Phys. Rev. B 68, 214415 (2003)
\bibitem{prb76.180407}Rajiv R. P. Singh and David A. Huse, Phys. Rev. B 76, 180407(R) (2007)
\bibitem{jap69.5962}J. B. Marston and C. Zeng, J. Appl. Phys. 69, 5962 (1991)
\bibitem{ap321.2}A. Kitaev, Ann. Phys. (N.Y.) 321, 2 (2006).
\bibitem{prl99.247203}H. Yao and S. A. Kivelson, Phys. Rev. Lett. 99, 247203 (2007).
\bibitem{prb81.104429}G. Kells, D. Mehta, J. K. Slingerland, and J. Vala, Phys. Rev. B 81, 104429 (2010).
\bibitem{prb62.R6065}K. Ohgushi, S. Murakami, and N. Nagaosa, Phys. Rev. B 62, R6065 (2000)
\bibitem{prb81.134418}B.-J. Yang, A. Paramekanti, Y.B. Kim, Phys. Rev. B 81, 134418(2010).
\bibitem{prb80.064404}T. P. Choy and Y. B. Kim, Phys. Rev. B 80, 064404 (2009).
\bibitem{prb81.205115}A. R\"uegg, J. Wen, and G. A. Fiete, Phys. Rev. B 81, 205115 (2010).
\bibitem{prb82.075125}Jun Wen, Andreas R\"uegg, Joseph C.-C. Wang and Gregory A. Fiete Phys. Rev. B 82, 075125 (2010)
\bibitem{prb74.205414}Y. Hatsugai, T. Fukui, and H. Aoki, Phys. Rev. B 74, 205414 (2006)
\bibitem{prb77.125119}Zhigang Wang and Ping Zhang Phys. Rev. B 77, 125119 (2008)

\bibitem{prb54.R17296}H. Aoki, M. Ando, and H. Matsumura, Phys. Rev. B 54, R17296 (1996).

\bibitem{Nagaosa}Nagaosa, Science 318, 758(2007).

\bibitem{acie46.6076}Y.-Z. Zheng, M.-L. Tong, W. Xue, W.-X. Zhang, X.-M. Chen, F. Grandjean, and G. J. Long, Angew. Chem., Int. Ed. 46, 6076(2007)
\end{thebibliography}
\end{document}